\documentclass{article}

\usepackage[
  top=2.5cm,
  bottom=2.5cm,
  left=2.5cm,
  right=2.5cm
]{geometry}
\usepackage{graphicx}%
\usepackage{multirow}%
\usepackage{multicol}%
\usepackage{amsmath,amssymb,amsfonts}%
\usepackage{amsthm}%
\usepackage{mathrsfs}%
\usepackage[title]{appendix}%
\usepackage{xcolor}%
\usepackage{textcomp}%
\usepackage{manyfoot}%
\usepackage{booktabs}%
\usepackage{algorithm}%
\usepackage{algorithmicx}%
\usepackage{algpseudocode}%
\usepackage{listings}%
\usepackage[font=small,labelfont=bf,justification=justified,singlelinecheck=false]{caption}
\usepackage{cite}  
\usepackage[colorlinks=true, linkcolor=blue, citecolor=blue, urlcolor=blue]{hyperref}

\theoremstyle{thmstyleone}%
\theoremstyle{thmstyletwo}%

\theoremstyle{thmstylethree}%

\title{Solution-derived barium titanate waveguides for integrated electro-optic modulation}

\author{%
\parbox{\textwidth}{\centering
Virginia Falcone$^{*}$, Eleni Prountzou, Jost Kellner,\\
Ülle-Linda Talts, Rachel Grange\\[0.6ex]
{\small Department of Physics, ETH Zurich, Institute for Quantum Electronics,\\
Optical Nanomaterial Group, Zurich, Switzerland}\\[0.6ex]
{\small $^{*}$Corresponding author: {vfalcone@phys.ethz.ch}}\\
}}

\date{}

\begin{document}
\maketitle


\abstract{Metal oxides with strong nonlinear optical properties and wide transparency window are key materials for the development of compact and efficient photonic integrated circuits used for electro-optic modulators and entangled photon sources. Among them, barium titanate (BaTiO$_{3}$) is particularly attractive due to its large Pockels coefficient. 
However, its use has been limited by challenges in material synthesis and in nanopatterning, owing to its chemical stability and inertness.

Here, we demonstrate a monolithic electro-optic modulator entirely based on solution-deposited BaTiO$_{3}$, fabricated through a bottom-up soft nanoimprinting lithography process. Fine-tuning the synthesis and nanofabrication enhances the optical properties of the polycrystalline material. By optimizing the process parameters, we achieve a reduction in propagation losses of two orders of magnitude, enabling efficient electro-optic modulation.
This scalable, etch-free approach enables direct patterning of high-quality BaTiO$_{3}$ structures, establishing a new route for low-cost, large-scale integrated electro-optic devices entirely based on oxide material compatible with a wide range of substrates.}
\maketitle

\section{Introduction}\label{sec1}

Recent advances in telecommunications have increased the demand for materials that enable compact device architectures, low power consumption, reduced thermal losses, and ultrafast operation. The transduction of signals from the electrical to optical domain is achieved by controlled alteration of the materials optical properties via harnessing the nonlinear Pockels effect in material systems. These materials, characterized by pronounced nonlinear effects, require a non-centrosymmetric crystal structure \cite{Wen}. In recent years, a significant branch of research has focused on non-centrosymmetric metal oxides, particularly lithium niobate (LiNbO$_{3}$, LNO) and barium titanate (BaTiO$_{3}$, BTO), owing to their strong nonlinear optical properties and broad transparency windows  \cite{Wen,Chen}.

One of the major challenges associated with these materials is their growth, which requires controlling both the elemental composition and the crystallographic phase, in addition to the challenges inherent to their fabrication.
Historically, they have been used in bulk crystal form, and only recently high-quality LNO thin films suitable for integrated photonics have become commercially available. These films, produced on various substrates through crystal-ion slicing and wafer bonding \cite{Rabiei}, have enabled the development of a wide range of LNO-based integrated devices, including electro-optic modulators \cite{Wang,Sabatti,Renaud,Xue}, optical parametric oscillators \cite{Kellner}, and photon pairs sources based on spontaneous parametric down-conversion \cite{Zhao,Harper}.
Compared with conventional CMOS materials such as Si or GaAs, nanopatterning of metal oxides presents several complications due to their chemical inertness and etching byproducts. These metal oxides are almost inert to standard wet and dry etching techniques. Therefore, device manufacturing typically relies on low-selectivity physical etching methods, such as inductively coupled plasma, followed by post-etch removal of redeposited material. This approach usually limits the achievable sidewall angle of fabricated photonic components to approximately 60° \cite{Kaufmann}. The use of hard masks, such as diamond-like carbon, can increase this value to around 80° \cite{Li}, but introduces additional fabrication complexity.

 Compared to LNO, BTO single crystals feature a significantly higher Pockels coefficient (r$_{42}$ = 1300 pm/V in the unclamped case \cite{Zgonik}) making it an attractive candidate especially for high-performance electro-optic applications. However, the development of commercially available BTO films is still in its early stages. The progress remains hindered by challenges related to crystal growth, material availability, and fabrication issues similar to those encountered for LNO \cite{Kim}.

BTO thin films are typically grown by techniques such as chemical vapour deposition, molecular beam epitaxy, pulsed laser deposition, radiofrequency sputtering or spalling from bulk substrates \cite{Kormondy,Lyu,Wagué,Mi,Thureja}. These approaches often constrain the choice of compatible substrates due to lattice mismatch and thermal expansion differences, or introduce limitations related to surface quality.
A common strategy is to employ crystalline seed layers, such as SrTiO$_3$, MgO, or DyScO$_3$ \cite{Meier,Posadas,Abel,Cao2}, that can be followed by a wafer-bonding step \cite{Ortmann,Abel2}. BTO thin films grown using these approaches have been employed as the electro-optic active layer underneath waveguides fabricated from other materials, such as Si or SiN \cite{Eltes,Abel2,Abel3,Tang,Eltes2,Mercier,Ummethala,Hiltunen,Ortmann,Xiong,Eltes3}. Furthermore, their refractive index variation under an applied electric field has been exploited to tune the optical mode confined within the waveguide \cite{Eltes,Abel2,Tang,Eltes2,Mercier,Ummethala,Hiltunen,Ortmann,Xiong}. This strategy has enabled the realization of hybrid electro-optic modulators that harness the nonlinear properties of BTO while employing different core materials, achieving favorable optical losses and high electro-optic efficiencies \cite{Ummethala,Abel2,Abel3,Mercier,Eltes,Xiong,Kohli}.
Monolithic electro-optic modulators based entirely on BTO have subsequently been developed using epitaxial thin films \cite{Dong,Lin,Petraru,Petraru2}, but they remain constrained by the complex growth and the top-down etching processes required for device fabrication \cite{Dong,Lin,Riedhauser,Kim}. 

In parallel, solution-based synthesis methods such as chemical solution deposition have emerged, offering greater flexibility in substrate selection \cite{Edmondson} and compatibility with large-area processing.
BTO films produced via the solution-based method exhibit a Pockels coefficient, approximately $25\, \mathrm{pm/V}$ \cite{Edmondson}, comparable to that of single-crystal LNO ($30\,\mathrm{pm/V}$ \cite{Wen}), making them highly promising for integrated electro-optic modulators.
Applications of solution-based BTO for electro-optic modulation, as uniform films beneath Si or SiN waveguides \cite{Picavet} or as electro-optic metasurfaces \cite{Weigand}, have been investigated; however, a fully monolithic integrated photonic circuit has not yet been realized.

\begin{figure}[h!]
  \centering
  \includegraphics[width=0.85\textwidth]{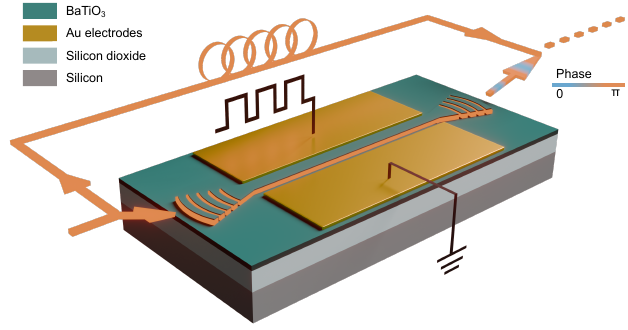}
  \captionsetup{width=0.9\textwidth}  
  \caption{\textbf{BTO sol-gel electro-optic modulator.} 
           Schematic of the monolithic electro-optic modulator based on sol-gel BTO. 
           The laser output is split into two arms: one is coupled into the BTO waveguide, 
           where an applied AC voltage induces a phase modulation between 0 and $\pi$, 
           while the reference arm remains unmodulated. Their recombination produces an amplitude-modulated output signal.}
  \label{fig:intro}
\end{figure}

Here, we present the first integrated electro-optic modulator entirely based on BTO sol-gel (Fig.\,\ref{fig:intro}\textcolor{blue}), fabricated using soft-nanoimprinting lithography (SNIL, details in Methods). This scalable bottom-up approach overcomes the challenges typically associated with metal oxide top-down patterning, enabling the definition of high-aspect-ratio structures with resolutions comparable to those achieved by electron-beam lithography \cite{Modaresialam,Verschuuren,Weigand,Talts,Talts1}.
Optical characterizations of the imprinted BTO waveguides, including propagation losses and electro-optic response, reveal both the intrinsic properties of this polycrystalline material and the strong tunability of device performance enabled by control of the sol-gel processing parameters.

Together, these results establish BTO sol-gel as a scalable, etchless platform for integrated photonic circuits based on non-centrosymmetric metal oxides, overcoming the substrate constraints and top-down fabrication limitations of conventional thin-film growth approaches.



\section{Thin film preparation and structural characterization}\label{sec2}

\begin{figure}[h!]
  \centering
  \includegraphics[width=0.95\textwidth]{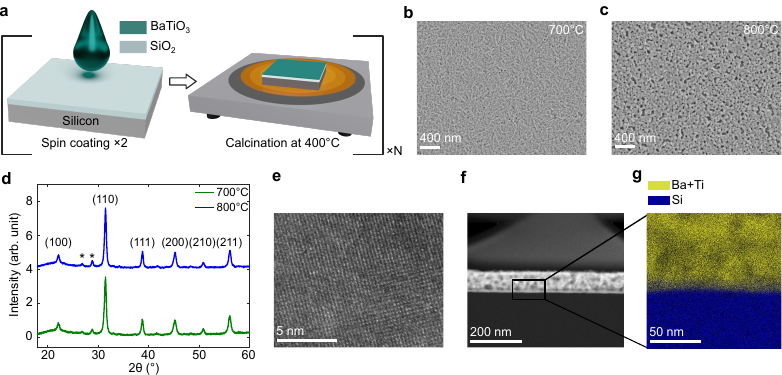}
  \captionsetup{width=0.9\textwidth}  
  \caption{\textbf{Material analysis of BTO sol-gel}. (a) Fabrication process of BTO sol–gel thin films. After spin-coating two BTO sol-gel layers, a calcination step at 400$^{\circ}$C is performed to remove organic by-products. This sequence is repeated N times to achieve the desired film thickness. SEM top-view images of two BTO sol-gel samples after final high-temperature annealing at (b) 700$^{\circ}$C and (c) 800$^{\circ}$C. The lower temperature film exhibits a more uniform morphology with smaller grains. (d) XRD patterns of the 700$^{\circ}$C (green) and 800$^{\circ}$C (blue) annealed samples. Peaks marked with asterisks are assigned to the Ba$_{2}$Ti$_{5}$O$_{12}$ phase. Reference XRD peak positions correspond to BaTiO$_{3}$ (PDF no. 00-005-0626) and to Ba$_{2}$Ti$_{5}$O$_{12}$ (PDF no. 00-017-0661). (e) HR-TEM of a single domain of the polycrystalline BTO sol-gel annealed at 800$^{\circ}$C. (f) STEM dark-field image of a 800$^{\circ}$C annealed BTO film cross-section on a SiO$_{2}$/Si substrate. (g) EDX elemental analysis  at the interface, black square in the STEM image, between the substrate and the BTO film. Si element is shown in blue (K$\alpha$ = 1.74 keV), while yellow represents the combined signal of Ba (L$\alpha$ = 4.46 keV) and Ti (K$\alpha$ = 4.51 keV) elements.}
 \label{fig:material}
\end{figure}

The sol-gel process enables the synthesis of inorganic oxide materials through the condensation and cross-linking of metal-organic precursors \cite{Edmondson}. The resulting amorphous matrix crystallizes into a polycrystalline structure after high-temperature annealing. The BTO sol-gel can be deposited on various substrates (MgO, fused quartz \cite{Weigand,Talts}, Si, SiO$_{2}$, SiN) by the spin-coating process, achieving films with thicknesses from tens to hundreds of nanometers. The thickness is defined by the spin-coating parameters (see Methods) and by the number of layers. For multilayer films, we identified a critical calcination step at 400$^{\,\circ}$C after every second layer to ensure the evaporation of organic sub-products, before depositing the following layers (Fig.\,\ref{fig:material}\textcolor{blue}{a}). Once the desired thickness is reached, a final high-temperature annealing is carried out to obtain polycrystalline BTO films.

The BTO grain size and degree of crystallinity are critically governed by the processing parameters, and need to be characterised and optimised to ensure the desired material properties. In this work, we investigated two different final annealing temperatures, 700$^{\circ}$C and 800$^{\circ}$C, which are commonly employed in BTO sol-gel processing \cite{Edmondson,Gust}, and compared the resulting polycrystalline BTO sol-gel structures in terms of both crystalline properties and device performance.
Two BTO sol-gel samples, each consisting of 12 layers for a total thickness of approximately 250 nm, were prepared on a SiO$_{2}$/Si substrate and were annealed in ambient atmosphere at 700$^{\circ}$C and 800$^{\circ}$C, with a ramp rate of 2$^{\circ}$C/min. 
The reduced ramp rate is required to mitigate the thermal stress between the BTO films and the substrate, arising from the high thermal expansion coefficient of this metal oxide \cite{Rhodes,Megaw,Bland}. Combined with lattice mismatch, these stresses promote cracking in thick films, effectively limiting the maximum achievable thickness for a BTO sol-gel with a molar concentration of 0.2 mol/L to approximately 500 nm.

The film morphology was first investigated by top-view scanning electron microscopy (SEM) images. Fig.\,\ref{fig:material}\textcolor{blue}{b,c} show the surface morphology after annealing at 700$^{\circ}$C and 800$^{\circ}$C. The film annealed at 800$^{\circ}$C exhibits increased grain size and higher porosity compared with the 700$^{\circ}$C sample, resulting in a less uniform surface.
The porosity of the BTO sol-gel films was estimated from refractive index measurements (Supplementary Fig.\,S2) using the Maxwell-Garnett effective medium approximation \cite{Markel}, assuming air-filled pores within the BTO layer. For the 800$^{\circ}$C-annealed sample, the porosity was estimated to be 30\%, whereas for the 700$^{\circ}$C-annealed film it decreased to 20\% (see Methods).

The structural properties were analyzed by X-ray diffraction (XRD) to assess the crystallinity of the films (Fig.\,\ref{fig:material}\textcolor{blue}{d}).
The diffraction patterns confirm the polycrystalline nature of both BTO films. Multiple diffraction peaks associated with different plan orientations, such as (100), (110), and (200), are observed, indicating a multi-domain structure without a preferential orientation. From the XRD peak broadening, accounting for both crystallite size and microstrain contributions, the grain size was estimated to be approximately 25 nm for the 700$^{\circ}$C film and 45 nm for the 800$^{\circ}$C sample. 
The broad diffraction peaks prevent a quantitative assessment of the tetragonal phase fraction in the sol-gel, typically indicated by the splitting of the (002) and (200) peaks \cite{Abel2,Pasuk}. To verify the presence of the tetragonal phase in the sol-gel BTO, Raman spectroscopy measurements were performed both on the 700$^{\circ}$C and 800$^{\circ}$C samples (Supplementary Fig.\,S1). In both cases, a distinct peak at 305 cm$^{-1}$ corresponding to the antisymmetric vibrational mode is observed, thereby confirming the presence of tetragonal phase in the BTO solution-derived films \cite{Vogler,Hayashi}. 

Transmission electron microscopy (TEM) was performed to further analyze the crystallinity of the BTO films and to investigate the grain structure observed in SEM. A high-resolution TEM (HR-TEM) image of a single crystal domain within the polycrystalline BTO film (Fig.\,\ref{fig:material}\textcolor{blue}{e}), confirmed the well-defined lattice fringes and thus the crystalline nature of the sol-gel derived BTO. Dark-field STEM imaging (Fig.\,\ref{fig:material}\textcolor{blue}{f}) shows the presence of a multidomains structure randomly orientated  and highlights the presence of partial porosity within the film. 
To better characterise the film quality and the interface with the substrate, energy-dispersive X-ray (EDX) analysis was carried out at the BTO-SiO$_{2}$/Si interface. EDX elemental mapping (Fig.\,\ref{fig:material}\textcolor{blue}{g}) shows the spatial distribution of Si (blue) and Ba/Ti (yellow). The yellow channel corresponds to the overlay of Ba (red) and Ti (green), revealing an approximately 10 nm interdiffusion region of Si at the BTO-SiO$_{2}$/Si interface. The EDX signal further confirms the presence of nanoscale porosity within the BTO film. This partial porosity, together with the grainy morphology, affects the optical and electrical properties of the film, as its refractive index and dielectric permittivity, respectively. These material characteristics lead to a slightly lower refractive index compared to single-crystal BTO \cite{Karvounis}. Ellipsometry measurements in the visible range, fitted using a Sellmeier dispersion model, yield an extrapolated refractive index of 2.03 and 1.92 at 1550 nm for 700$^{\circ}$C and 800$^{\circ}$C samples, respectively (Supplementary Fig.\,S2). The 800$^{\circ}$C film exhibits higher porosity than the 700 °C sample, leading to a correspondingly lower effective refractive index.

The relative high refractive indeces allow contrast between the BTO sol-gel and the surrounding media enabling efficient optical confinement. This property combined with the intrinsic Pockels effect of BTO, provides the fundamental ingredients for demonstrating an electro-optical modulator fully based on this class of material. The material analysis by XRD and TEM measurements confirmed the polycrystalline structure of the BTO sol-gel and the absence of a preferential orientation in the deposited films. Although XRD and Raman patterns for the samples annealed at 700$^{\circ}$C and 800$^{\circ}$C appear nearly identical, pronounced morphological and compositional differences are observed, motivating a comparative analysis of their impact on device performance.

\section{Integrated electro-optic modulator}\label{sec3}

\begin{figure}[h!]
\begin{center}
  \includegraphics[width=0.85\textwidth]{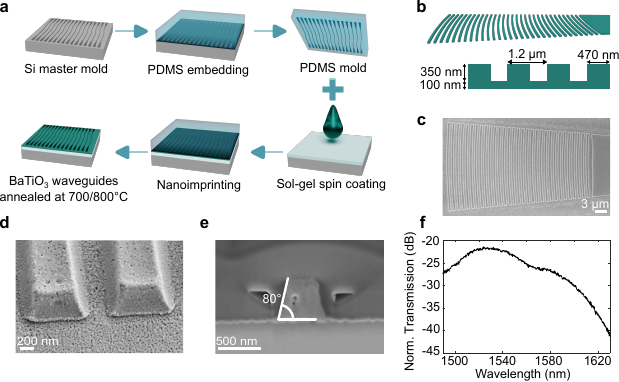}
  \captionsetup{width=0.9\textwidth}  
\caption[Fabrication of BTO sol-gel waveguide]{ \textbf{Fabrication of BTO sol-gel waveguides}. (a) Sketch of the soft nanoimprint lithography (SNIL) process used to fabricate the BTO sol-gel waveguides as described in the text. (b) Schematic of the simulated grating couplers used to optimise coupling with the BTO sol-gel waveguide. The optimized grating coupler parameters are the following: width = 470 nm, height = 350 nm and a period = 1.2 $\mu\text{m}$. (c) SEM top-view image of the fabricated grating couplers achieved by the bottom-up approach shown in (a). (d) SEM image taken at a 30° tilt showing a higher-magnification view of the grating sidewalls. (e) SEM cross-section image of the grating coupler obtained by focused ion beam milling, confirming a uniform filling of the PDMS mold by the BTO sol-gel. (f) Normalized transmission spectrum of a waveguide of length $400\,\mu\text{m}$ with input and output grating couplers.}
\label{fig:fabgc}
\end{center}
\end{figure}

The use of a BTO sol-gel enables a new route for the fabrication of waveguides by soft-nanoimprinting lithography. This scalable bottom-up process, based on the imprinting of the geometry directly into the material while it is still in its liquid state, overcomes the etching challenges associated with the chemical inertness of this metal oxide. Fig.\,\ref{fig:fabgc}\textcolor{blue}{a} shows a sketch of the entire fabrication process. A silicon master mold, with a nanostructured design, is produced via electron beam lithography and reactive ion etching. To compensate for the approximately 50\% shrinkage occurring in all dimensions during sol-gel annealing, the master mold features are scaled accordingly.
A flexible polydimethylsiloxane (PDMS) stamp is then generated by embedding the silicon master mold (see Methods for details) \cite{Talts}. Immediately after sol-gel spin coating on the SiO$_{2}$/Si substrate, the PDMS mold is used to imprint the design, which upon final annealing yields a polycrystalline BTO sol-gel waveguide.

In this work, the edge couplers, typically reported for BTO waveguides \cite{Lin,Petraru,Abel}, are replaced by grating couplers.  Finite-difference time-domain simulations were carried out to design the BTO sol-gel grating couplers, taking into account the refractive index of the material and fabrication shrinkage limitations \cite{Modaresialam}. In particular, shrinkage compensation in the silicon master mold limits the grating fill factor to values below 50\%. Moreover, the combined thickness of the imprinted structures and the residual BTO layer should not exceed ~500 nm to prevent crack formation in the sol-gel film. Considering these limitations, the optimized grating couplers, shown in Fig.\,\ref{fig:fabgc}\textcolor{blue}{b}, consist of elements with period of 1.2 $\mu\text{m}$, width of 470 nm and height of 350 nm in addition to 100-nm BTO residual layer. This design provides high directional emission efficiency of the grating coupler across the analyzed wavelength range and yields theoretical coupling losses of 4 dB per grating (Supplementary Fig.\,S3). A top-view SEM image of a fabricated grating coupler is shown in Fig.\,\ref{fig:fabgc}\textcolor{blue}{c}. The resulting sidewalls exhibit angles of $\approx$ 80$^{\circ}$, as visible in the SEM image in Fig.\,\ref{fig:fabgc}\textcolor{blue}{d} and confirmed by cross-sectional imaging in Fig.\,\ref{fig:fabgc}\textcolor{blue}{e}. The latter further demonstrates the uniform filling of the BTO sol-gel within the grating structures after the nanoimprinting. This represents the first demonstration of a grating coupler fabricated by a cost-effective and scalable process such as SNIL for this class of materials. The implementation of this bottom-up process enables higher sidewall angles compared to conventional top-down etching of metal oxides, which is typically limited by redeposition effects \cite{Kaufmann}.

\begin{figure}[h!]
  \centering
  \includegraphics[width=0.95\textwidth]{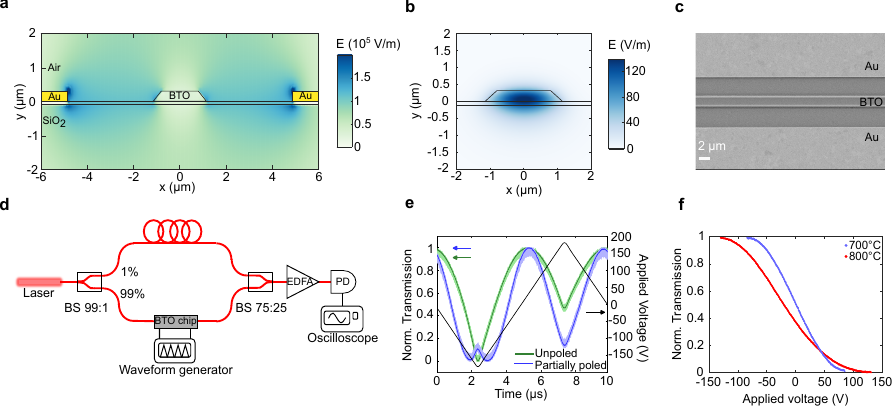}
  \captionsetup{width=0.9\textwidth}  
\caption[Material analysis of BTO sol-gel]{ \textbf{Electro-optic characterization}. FEM simulations of the electric (a) and optical (b) field distributions for a BTO sol-gel waveguide on a 100 nm residual BTO layer on a SiO$_2$/Si substrate. The two top electrodes are separated by $9.7  \mu\text{m}$. The electrostatic simulation is performed with an applied voltage of 1 V. (c) SEM image of the fabricated electrodes close to the waveguide. (d) Sketch of the experimental setup used to achieve interference between the modulated light transmitted through the BTO waveguide and a reference fiber arm outside the chip. (e) Normalized transmission signal as a function of time for the 700$^{\circ}$C sample before (solid green line) and after (solid blue line) DC poling. The application of a DC bias enables domain alignment, resulting in a $\pi$-phase modulation at a driven frequency of 100 kHz. (f) Normalized transmission signal as a function of the applied voltage. The sample annealed at 700$^{\circ}$C (blue line) exhibits a lower $V_{\pi} L$ compared to the 800$^{\circ}$C-annealed sample (green line).}
\label{fig:meas}
\end{figure}

The normalized transmission spectrum of a $400\,\mu\text{m}$-long waveguide, including both input and output grating couplers, measured around 1550\,nm, is shown in Fig.\,\ref{fig:fabgc}\textcolor{blue}{f}, with a measured coupling losses of about 10 dB per grating. This value, higher than predicted by simulations, arises from small deviations in the final grating dimensions caused by sol-gel shrinkage during annealing.
We simulated both the electric field distribution and the optical mode of the waveguide (Fig.\,\ref{fig:meas}\textcolor{blue}{a,b}) with finite element method (FEM), enabling optimization of the device geometry for efficient electro-optic performance. The waveguide core was set to a width of $2\,\mu\text{m}$  to achieve the optical field distribution shown in Fig.\,\ref{fig:meas}\textcolor{blue}{b}, confirming mode confinement.

A  100\,nm-thick BaTiO$_{3}$ film (four layers) was first deposited on a SiO$_2$/Si substrate via the previously described spin-coating process, ensuring the residual BTO thickness beneath the waveguides matched the optimal value predicted by grating coupler simulations. On this film, single-arm modulators of different lengths, consisting of 350\,nm-thick s-bend section followed by a straight segment, with grating couplers of the dimensions shown in Fig.\,\ref{fig:fabgc}\textcolor{blue}{b}, were realized using SNIL.
The fabrication of BTO sol-gel waveguides with lengths ranging from 100 to 2000 $\mu\text{m}$ enabled the estimation of propagation loss.
The sample annealed at a temperature of 800$^{\circ}$C exhibited propagation loss of approximately 50 dB/cm, whereas reducing the final annealing temperature to 700$^{\circ}$C decreased the loss to $\sim\,$20 dB/cm (Supplementary Note 4). The estimated losses were obtained by fitting the transmitted light through waveguides of different lengths at a wavelength of 1550 nm, providing the first quantitative estimate of optical losses in BTO sol-gel.
The reduction of the loss by two orders of magnitude for the 700$^{\circ}$C sample is attributed to its different morphology, as shown in Fig.\,\ref{fig:material}\textcolor{blue}{b}. Compared to the 800$^{\circ}$C sample, the lower-temperature film is smoother and shows reduced porosity, resulting in decreased scattering losses.
These results demonstrate the significant performance gains enabled by targeted optimization of the sol-gel processing steps. Adjusting these parameters directly impacts the material properties and, consequently, the optical performance of the final device.

Two top electrodes, Cr (5 nm)/Au (300 nm), were patterned on either side of the waveguides by optical lithography, followed by electron-beam evaporation and lift-off (Fig.\,\ref{fig:meas}\textcolor{blue}{c}). The simulated electric field distribution between the fabricated top electrodes, separated by $9.7\,\mu\text{m}$ (Fig.\,\ref{fig:meas}\textcolor{blue}{a}), was obtained using a relative dielectric permittivity of 3.9 for the SiO$_2$ and of 500 for the BTO sol-gel \cite{Edmondson}.
Owing to this high permittivity, a significant fraction of the applied electric field is displaced into the surrounding lower-permittivity materials, namely the air above and SiO$_2$ below the structure. Indeed, FEM simulation predicts, under an applied voltage of 1 V, an average electric field of $5.1 \times 10^{4}$ V/m in the 100 nm residual BTO layer below the waveguide and a mean value of $3.3 \times 10^{4}$ V/m in the waveguide core.
The electrodes extend along the full length of each s-bend waveguide, maximizing the active length of the BTO sol-gel. This is enabled by the polycrystalline nature of the BTO sol-gel, which exhibits no preferred crystallographic orientation, as confirmed by the XRD patterns (Fig.\,\ref{fig:material}\textcolor{blue}{d}). Consequently, the electro-optic response is uniform along the entire waveguide, with no directional dependence on the applied voltage across the electrodes. This uniformity contrasts with monocrystalline films, whose response is intrinsically anisotropic due to their fixed crystallographic orientation. 

The amplitude modulation was measured using the optical setup shown in Fig.\,\ref{fig:meas}\textcolor{blue}{d}.
A continuous-wave laser source at 1550 nm was split, with 99\% of the power coupled into the chip and the remaining 1\% directed into a reference fiber arm. The optical signal propagating through the BTO waveguide experiences a phase modulation induced by the applied electric field. The reference arm was matched in length to the on-chip path, and optical attenuators were used to balance its power with the transmitted power from the output waveguide. Upon off-chip recombination, the two optical signals interfere, yielding an amplitude modulated signal that was amplified by an erbium-doped fiber amplifier (EDFA) and detected using an InGaAs fiber-optic receiver. 
Electro-optic modulation was driven by a triangular waveform, generated by a function generator and amplified to a peak-to-peak voltage of V$_\text{pp}$ = 360 V, at a frequency of 100 kHz.
The modulation properties of this low-cost material platform were investigated by characterising BTO sol-gel waveguides of different lengths, in samples annealed at 700 $^{\circ}$C and 800 $^{\circ}$C. The modulated response obtained for a 2000 $\mu\text{m}$ waveguide is reported in Fig.\,\ref{fig:meas}\textcolor{blue}{e} for the 700 $^{\circ}$C sample. The as-deposited BTO sol-gel modulator was not able to induce a $\pi$-phase modulation, as illustrated by the green line in Fig.\,\ref{fig:meas}\textcolor{blue}{e}, since the predominantly random orientation of the domains partially suppresses the effective electro-optic response \cite{Dong,Abel}.
To overcome this limitation, a DC bias of 60 V was applied across the top electrodes to align the domains. Given the initially random domain orientation, the applied electric field partially aligns the domains toward the field direction, resulting in a partially poled film.
After this poling step, the DC bias was turned off and AC modulation was activated. The resulting blue line in Fig.\,\ref{fig:meas}\textcolor{blue}{e} shows that a full $\pi$-phase modulation is achieved after partial poling, demonstrating the operation of a partially poled BTO sol-gel integrated modulator and allowing extraction of the half-wave voltage required to reach this $\pi$ phase shift ($V_{\pi}$) for the BTO sol-gel. A comparable result was achieved for the 800$^{\circ}$C sample following the application of a DC voltage (Supplementary Fig.\,S5). Owing to the different morphologies, the two annealed films require distinct DC electric field values for achieving a $\pi$-phase modulation: for the 700$^{\circ}$C-annealed sample, a field of $65\, \mathrm{kV/cm}$ is sufficient, whereas the 800$^{\circ}$C-annealed film requires almost three times this value ($190\, \mathrm{kV/cm}$), reflecting the stronger pinning of domains in its coarse-grained morphology.

The polycrystalline BTO morphology strongly influences the modulation efficiency of the devices. Figure\,\ref{fig:meas}\textcolor{blue}{f} shows the modulated signal as a function of the applied voltage for the 800$^{\circ}$C and 700$^{\circ}$C samples, shown in red and blue, respectively. The figure of merit $V_{\pi}\,L$, which is independent of device length, was estimated from multiple measurements on waveguides of different lengths for both samples. 
Values of $52 \pm 2\,\mathrm{V{\cdot}cm}$ and $35 \pm 2\,\mathrm{V{\cdot}cm}$ were obtained for the 800$^{\circ}$C and 700$^{\circ}$C-annealed samples, respectively.
Thanks to its higher refractive index, resulting from the lower porosity of the film, the 700$^{\circ}$C sample provides stronger optical confinement of the guided mode, which in turn contributes to its higher modulation efficiency. Moreover, the smaller grains allow the domains to be poled with a lower applied DC field, enabling a higher modulation efficiency compared to the 800$^{\circ}$C-annealed sample. This demonstrates that optimizing the BTO sol-gel properties and its morphology significantly reduces $V_{\pi}\,L$, enabling efficient, compact, and easily fabricated modulators, and highlighting the potential for further performance improvements.

It is important to note that these $V_{\pi}\,L$ values arise from the high dielectric permittivity of the BTO sol-gel \cite{Edmondson} that limit the electric field magnitude within the BTO waveguide and residual layer compared to the surrounding air and SiO$_{2}$ (Fig.\,\ref{fig:meas}\textcolor{blue}{a}). From FEM simulations, the overlap integral $\Gamma$ \cite{Ghione} between the optical field (Fig.\,\ref{fig:meas}\textcolor{blue}{b}) and the electric field distribution (Fig.\,\ref{fig:meas}\textcolor{blue}{a}) in the waveguide and residual layer was calculated, yielding a value of $\sim\,20\%$. Using this overlap value and the experimentally measured $V_{\pi}\,L$ of $35 \pm 2 \mathrm{V{\cdot}cm}$, the effective electro-optic coefficient $r_\text{eff}$ of the BTO sol-gel can be estimated from the relation \cite{Ghione}:
\[
r_\text{eff} = \frac{\lambda_0\,G\,n_\text{eff,m}}{V_{\pi}\,L\,n_\text{eff}^4\,\Gamma}
\]
where ${\lambda_0}$ is the operating wavelength (1550 nm), ${n_\text{eff,m}}$ is the effetive refractive index of the optical mode, ${n_\text{eff}}$ is the effective refractive index of the BTO, and $G$ is the gap between the top electrodes.
The estimated value of $r_\text{eff}$ of approximately $20\, \mathrm{pm/V}$ for the poled 700$^{\circ}$C sample is in line with previously reported values \cite{Edmondson,Weigand}, supporting the electro-optic response of the sol-gel BTO. 
Although this value is one order of magnitude lower than that of bulk BTO, it approaches the typical electro-optic coefficient of LiNbO$_{3}$ \cite{Wen}. This demonstrates that a polycrystalline sol-gel BTO is a suitable platform for electro-optic modulators fabricated through scalable soft nanoimprinting lithography.
Beyond the performance demonstrated in this work, further optimization of film crystallinity and domain configuration offers a clear route toward lower $V_{\pi}\,L$ and reduced optical losses. Depositing BTO sol-gel on thin seed layers, such as La$_2$O$_2$CO$_3$, represents a promising strategy to improve crystallinity and reduce film porosity. Reducing lattice mismatch with the substrate can promote the formation of in-plane oriented tetragonal phase, which further enhance the electro-optic modulation efficiency of sol-gel BTO \cite{Picavet}. Moreover, post-annealing treatments to reduce oxygen vacancies and leakage currents \cite{Choi}, as well as controlled domain orientation via temperature or optical-assisted poling \cite{Zhou,Sarott}, offer additional routes to further enhance electro-optic performance of this class of devices.

\section{Conclusion}\label{sec5}

In this work, we establish a new material platform based on polycrystalline BTO sol-gel for monolithic integrated photonic circuits, demonstrating its suitability for efficient electro-optic modulation and the tunability of device performance through processing parameters.
The sol-gel approach enables a scalable bottom-up fabrication route through SNIL, overcoming the challenges associated with conventional top-down processing typically required for monocrystalline material such as lithium niobate. Avoiding standard top-down etching techniques yields higher sidewall angle and mitigates redeposition-related limitations. Moreover, the chemical synthesis of BTO is not constrained by substrate choice, ensuring compatibility with silicon and dielectric photonic platforms and wafer-level manufacturing.

By correlating optical losses, electro-optic response, morphological and structural properties, we show that sol-gel processing parameters, such as annealing temperature, provide clear routes for performance optimization. Reducing the final annealing temperature from 800$^{\circ}$C to 700$^{\circ}$C decreases the porosity of the BTO waveguides, leading to an improvement in the optical properties of the polycrystalline material. This temperature reduction results in a two order of magnitude decrease in propagation losses, enabling efficient electro-optic modulation. The polycrystalline nature of solution-derived material, with its initially random domain orientation, allows the entire s-bend waveguide to contribute as an active modulation region, without being limited by intrinsic crystallographic alignment. At the same time, the relatively large Pockels coefficient of BTO enables compact modulators in which the short interaction length mitigates the impact of propagation losses. By implementing a poling process, partial domain alignment can be induced in integrated devices, enabling the achievement of full $\pi$-phase modulation in compact devices.

These results establish solution-processed BTO imprinted nanostructures as a tunable platform for integrated electro-optic modulation, paving the way for scalable metal oxides photonics in next-generation integrated circuits. The demonstrated phase and amplitude modulation capabilities suggest this platform as a promising building block for reconfigurable photonic processors, on-chip optical interconnects and emerging neuromorphic systems.


\section{Methods}\label{sec6}

\noindent\textbf{Material characterization}\\
XRD patterns and HR-TEM were performed to assess the polycrystalline structure of a 0.2 mol/L sol-gel derived BTO. XRD measurements were conducted using a PANalytical X’Pert PRO MRD diffractometer with a step size of 0.5$^\circ$/s, using Cu K$\alpha_{1}$ radiation ($\lambda = 1.5406$ Å) selected by a Ge monochromator. The samples analyzed are made of 12 layers corresponding to 250 nm thickness. Each layer is obtained by spin-coating 40 $\mu$L of sol-gel at 3500 rpm for 40 s, followed by 5 min of gelation in ambient air. After every two layers, an additional calcination step of 5 min at 400$^{\circ}$C was required. After reaching the desired thickness, a final furnace annealing at 700$^{\circ}$C or 800$^{\circ}$C for 2 h, with a ramp rate of 2$^{\circ}$C/min, was performed to achieve the sol-gel crystallization.

HR-TEM analysis was performed using a JEOL JEM F200 microscope. The lamellae, with a thickness $\approx$ 50\,nm, was prepared from the samples annealed at 800$^{\circ}$C.

\noindent\textbf{Soft nanoimprinting process}\\
For the soft nanoimprinting process, hybrid PDMS molds were prepared as described in a previous publication \cite{Talts}. First, a layer of hard-PDMS (h-PDMS) was spin-coated onto the silicon master mold at 1000 rpm for 40 s with an acceleration of 1000 rpm/s, followed by a curing step on a hot plate at 70$^{\circ}$C for 10 min. The mold with the cured h-PDMS layer was then placed in a beaker and covered with a layer of soft-PDMS (s-PDMS) to a thickness of approximately 4 mm.
The final curing was performed in an oven at 65$^{\circ}$C for a minimum of 12 hours.
Particular attention was paid to preventing bubble formation, caused by trapped gas within the high viscosity liquid, with any residual bubbles removed by degassing.
After curing, the PDMS mold was peeled from the Si mold and degassed at 200 mbar for 15 min prior to use in the nanoimprinting process.

The devices presented in this work were fabricated by imprinting 50 $\mu$L of a 0.2 mol/L BTO sol-gel spin-coated at 1000 rpm for 5 s onto a SiO$_{2}$/Si substrate using the PDMS molds obtained as described above.
Following imprinting, a curing step was carried out on a hot plate at 70$^{\circ}$C for 2 h before the separation of the PDMS mold from the imprinted BTO film, in order to remove part of the organic by-products. Imprinted chips were then annealed at 400$^{\circ}$C for 10 minutes on a hot plate to eliminate the remaining organics, followed by a final furnace annealing at 700$^{\circ}$C or 800$^{\circ}$C for 2 h with a ramp rate of 2$^{\circ}$C/min to achieve the sol-gel crystallization.

\noindent\textbf{Simulations}\\
Finite-element simulations were performed using COMSOL Multiphysics (version 6.2). The simulated geometry reproduced the final device stack: 100-nm BTO sol-gel residual layer, 350-nm imprinted waveguide on a SiO$_{2}$(2$\,\mu$m)/Si substrate. The refractive index of BTO was extrapolated at 1550\,nm from ellipsometry measurements in the visible range, fitted using a Sellmeier dispersion model.  A refractive index of 2.03 and 1.92 at 1550 nm for 700$^{\circ}$C and 800$^{\circ}$C samples, was estimated respectively (Supplementary Fig. S2). The electrostatics module was used to simulate the electric field distribution generated by two top electrodes separated by 9.7$\,\mu$m, using a relative dielectric
permittivity of 500 for the BTO sol-gel and of 3.9 for the SiO$_{2}$. The Wave optics module was used to simulate the optical mode profile of the BTO sol-gel waveguide. An extra-fine physics-controlled mesh was used in the simulations.

\noindent\textbf{Porosity estimation}\\
The Maxwell-Garnett effective medium approximation was used to estimate the porosity of the BTO films annealed at 700$^{\circ}$C and 800$^{\circ}$C, assuming air-filled pores ($n = 1$). 
For these calculations a refractive index of 2.3 was used for the bulk BTO \cite{Wemple}, while refractive indices of 2.03 and 1.92 at 1550\,nm was used for the films annealed at 700$^{\circ}$C and 800$^{\circ}$C, respectively (determined from ellipsometry measurements shown in Supplementary Fig.S2).

\noindent\textbf{Electro-optic measurements}\\
Electro-optic measurements were performed using a continuous-wave laser at 1550\,nm with an output power of 60\,mW. The laser is split using a 99:1 fused-fiber polarization splitter, such that the 99\% of the optical power is coupled into the device and the 1\% into a reference fiber arm. The chip arm passed through a fiber polarization controller and was coupled into the device by grating couplers at 10° incident angle. The reference arm was matched in length and power, by the use of attenuators, to ensure interference with the chip arm. The two optical signals were then recombined using a 75:25 fused-fiber polarization splitter, and the resulting signal was amplified by an erbium-doped fiber amplifier (EDFA). Detection was performed with an InGaAs fiber-optic receiver. A triangular AC drive voltage, V$_\text{pp}$ = 360 V at a frequency of 100\,kHz, was applied to the two top electrodes electrical contacted by micromanipulated probe needles. The modulated optical output and the electrical reference waveform were recorded on a digital oscilloscope.

\section*{Acknowledgements}

The authors thank the Scientific Center for Optical and Electron Microscopy (ScopeM), the Binning and Rohrer Nanotechnology Center (BRNC), and the FIRST cleanroom of ETH Zurich, as well as Prakhar Jain and Alessandra Sabatti for discussions related to fabrication. This work was supported by the Swiss National Science Foundation SNSF (Consolidator Grant No. 213713) and the Sinergia Program (Project No. CRSII5 206008).

\section*{Author contributions}

V.F. and R.G. conceived the study. V.F. and E.P. performed the material characterization, with the contribution from Ü.-L.T. V.F. conducted the experiments and simulations, analysed the data and wrote the original draft of the manuscript. E.P. synthesized the material and contributed to data analysis. J.K. developed the experimental setup and contributed to the electro-optic measurements, simulations and data analysis. V.F., Ü.-L.T. and E.P. fabricated the devices. R.G. provided funding and supervised the project. All authors discussed the results and approved the final manuscript.

\section*{Competing interests}

The authors declare no competing interests.

\section*{Data availability}

Data supporting the findings of this study are available within the Article and its Supplementary Information. Raw data are available from the corresponding author upon reasonable request.

\bibliographystyle{unsrt}
\bibliography{sn-bibliography}
\end{document}